# Signal Development for Saturated Ultrafast Sensors with Impact Ionization Gain

**Bruce A. Schumm**[*]

*Santa Cruz Institute for Particle Physics and Department of Physics, University of California at Santa Cruz, 1156 High Street, Santa Cruz, California, 95064, USA*
*E-mail*: `baschumm@ucsc.edu`

ABSTRACT: A closed-form approximate expression is presented for the short time-frame development of silicon diode sensor signals in the context of high frame-rate detection of incident X-ray fluxes, in the limit that the X-ray absorption profile generates a longitudinally-uniform distribution of electron-hole pairs in the detector bulk. The expression represents the immediate time development of signals from diode sensors both with (LGAD) and without (PIN) gain, and presents a temporal scale associated with the onset of gain. Principles limiting the detection frame rate in the presence of electronic readout noise are discussed. Making use of an elemental simulation, the relative advantage of LGAD vs. PIN diode sensors is explored as a function of the effective electronic collection time. It is found that for an idealized LGAD sensor with a gain of 30, the gain provided by impact ionization yields an advantage relative to PIN diode sensors for frame rates as high as 10 GHz.

---

[*] Corresponding author.

# Contents

## 1. Introduction

The development of next generation synchrotron and FEL X-ray sources raises the possibility of dynamic X-ray diffraction and imaging with multi-GHz frame rates [1,2]. Attendant to the development of this capability will be the need to produce high-dynamic-range accelerator diagnostics that can characterize both primary and secondary beams at a commensurate frame rate.

This paper reports on an exploration of the nature of signal development within silicon diode sensors, both with (LGAD) and without (PIN) impact-ionization gain, in concert with idealized fast-shaping readout electronics, that can shed light on the achievable frame-rate of solid-state sensors. Studies are performed in the "saturated" limit, which for the purpose of this paper is construed to mean that the deposition process creates an instantaneous, longitudinally-uniform distribution of electron-hole pairs that then undergo motion as they are collected on the electrodes of the reverse-biased diode sensor. Such a profile would arise in the case that the deposition is induced by a large number of quanta within a flux of particles arising from a beam pulse occupying a single RF bucket, which is the case envisioned for the next generation of high-current, high-flux light source facilities. However, in the development that follows, it is also assumed that the space-charge field that arises during the collection of these electron-hole pairs remains small relative to the field created by the reverse bias, so that the drift speed of both electrons and holes remains close to its saturated value during the collection process. Here, we consider depositions induced by a mono-energetic X-ray field, but the generalization to white beam, or a stream of ionizing particles, is straightforward. An approximate closed-form treatment of the signal development immediately after the instantaneous deposition event – the time frame relevant to high repetition rate applications – is presented, followed by an elemental Monte Carlo simulation. Each of these sheds somewhat independent light on the potential application of solid-state sensors to ultra-fast X-ray detection.

## 2. Closed-Form Approximate Treatment

For a deposition of the nature described above, involving the instantaneous absorption of many quanta, the deposition creates a plasma of electron-hole pairs, uniform in the depth of the detector, of density

$$\rho_0(\vec{x}) = \frac{E_\gamma \sigma(x,y)}{3.62\ \lambda}$$

(1)

where σ(x,y) is the transverse profile of the instantaneous incident flux in quanta per cm$^2$, $E_\gamma$ is the X-ray energy in eV, and λ is the attenuation length of the quantum in the material, here assumed to be large relative to the thickness of the sensor. The value 3.62 represents the mean amount of energy, in eV, needed to excite an electron into the conduction band in the silicon bulk [3]. In the parallel plate approximation, the charge collection rate will just be given by the fractional rate of motion through the sensor [4]; in the limit of no gain, such as that expected for a conventional PIN diode, this is given by the sum of the electron and hole contributions:

$$dQ(t) = \left[\sum_{e/h} \int d\vec{x} \frac{v_{e/h}(z)}{d} \rho_0(\vec{x})\right] dt$$

(2)

where Q is the amount of charge induced through the amplifier through the motion of the electrons or holes, d is the sensor thickness and $v_{e/h}(z)$ the local electron/hole drift velocity. Thus, under the assumption that the drift velocity is saturated throughout the bulk, so that $v_{e/h}(z) = v^s_{e/h}$ is independent of z, the charge collection rate immediately following the creation of the plasma is given by

$$dQ(t) = \frac{E_\gamma}{3.62 \, \lambda d} \sum_{e/h} v^s_{e/h} \iiint dxdydz\, \sigma(x,y)\, dt = \sum_{e/h} \frac{\Phi E_\gamma v^s_{e/h}}{3.62 \, \lambda}$$

(3)

where the integrated flux Φ, in number of quanta, is defined according to

$$\Phi \equiv \iint dxdy\, \sigma(x,y).$$

(4)

Gain is introduced via the impact ionization coefficient $\alpha_{e/h}$ which is the number of electron-hole pairs created per cm of travel of an extant carrier (electron or hole), and depends upon z through the sensor doping profile, as well as environmental parameters such as the ambient electric field and temperature. This leads to an increase in the plasma density proportional to the density itself, and to the path length $dl$ traversed in the time interval dt by the carriers in the region of the space point *x*:

$$d\rho_e(\vec{x},t) = d\rho_h(\vec{x},t) = \sum_{e/h} \alpha_{e/h}(z)\rho_{e/h}(\vec{x},t)dl_{e/h}(\vec{x}) = \sum_{e/h} \alpha_{e/h}(z)\rho_{e/h}(\vec{x},t)v^s_{e/h}dt$$

(5)

where $\rho_{e/h}(\vec{x},t)$ is the local electron or hole density, with $\rho_e(\vec{x},0) = \rho_h(\vec{x},0) = \rho_0(\vec{x})$. Including gain, the dependence of the carrier density upon time immediately following the creation of the plasma at t=0 will be dominated by the leading order term in the multiplication of the plasma:

$$\rho_{e/h}(\vec{x},t) \cong \rho_0(\vec{x})\left[1 + \sum_{e/h} \alpha_{e/h}(z)\text{sgn}_{e/h} v^s_{e/h} t\right]$$

(6)

where t is the time elapsed since the creation of the plasma (note that this approximation also assumes that the range of transport of the carriers over the elapsed time *t* is small relative to the scale of variation in $\alpha(z)$; neither of these approximations were made in the simulation described in the following section). The directional factor $\text{sgn}_{e/h}$ introduces a sign that is of opposite value for electrons and holes, to represent the opposing motion of the electrons and holes in the field of the silicon bulk. The introduction of gain into Equation (2), in this approximation, then leads to a charge collection rate of the form

$$dQ_{e/h} = \frac{\Phi E_\gamma}{3.62\, \lambda d} \int dz\, v^s_{e/h} \left[1 + \sum_{e/h} \alpha_{e/h}(z) v^s_{e/h} t\right] dt.$$

(7)

For LGAD sensors operating below their breakdown point, the gain due to hole-induced impact ionization is very small, and so $\alpha_h(z) \cong 0$. In this further approximation,

$$dQ_{e/h} = \frac{\Phi E_\gamma}{3.62\, \lambda} v^s_{e/h} [1 + A_e v^s_e t] dt$$

(8)

where the scaled impact ionization factor $A_e$ is given by

$$A_e \equiv \frac{1}{d} \int dz\, \alpha_e(z)$$

(9)

Integrating this over time, the collected charge at a short time t after the creation of the plasma is then, at leading order

$$Q_{e/h}(t) = \frac{\Phi E_\gamma v^s_{e/h}}{3.62\, \lambda} \left[t + \frac{1}{2} A_e v^s_e t^2\right].$$

(10)

In the limit of zero gain (PIN diode), this is apparently independent of sensor thickness d. However, it must be recognized that Equation (10) applies only for the first instant after the creation of the plasma. Over the full collection period charges reach the boundary of the bulk and stop contributing to both multiplication and charge collection. For example, for the PIN case, this results in the collection current immediately reaching a maximal value, which is that given by the constant (first) term in Equation (8), and then falling off with time.

When read out with electronics that collects charge for some characteristic time τ after the creation of the plasma, the total collected charge, related in this case to the peak height of the resulting signal pulse, is then, in this approximation,

$$Q_{e/h} = \frac{K \Phi E_\gamma v^s_{e/h} \tau}{3.62\, \lambda} \left[1 + \frac{1}{2} K A_e v^s_e \tau\right]$$

(11)

where *K* is a dimensionless constant of order 1 that relates the shaping time to the effective charge collection time and is dependent upon the particular type of shaping done by the amplifier. Again, τ is the shaping time, Φ is the total X-ray flux through the pixel in number of quanta, λ is the X-ray attenuation length, $E_\gamma$ is in eV, $v^s_{e/h}$ is the saturated drift velocity, and $A_e$ a scaled gain factor arising from impact ionization that depends upon the sensor bias and doping profile, as well as other environmental factors. The term quadratic in τ (the second term in brackets) is that due to the LGAD gain induced by the impact ionization process.

When applied to the detection of an instantaneous flux of X-rays, the minimal detectable signal of interest $Q_{min}$ must be larger than the readout noise by some factor (for self-triggered systems, this factor is typically of order 10, but it can be smaller for precisely gated systems). It has been generally argued [5] that as $\tau \to 0$ (the high frame-rate limit), series noise contributions will dominate, and grow with falling shaping time as $1/\sqrt{\tau}$. At the same time, Equation (11) suggests that the collected charge falls at least linearly with $\tau$. As a result, $Q_{min}$ will worsen (increase) as at least the 3/2 power of the frame rate. Thus, the achievable frame rate can depend strongly upon the total integrated signal, which can be enhanced by the impact ionization process via the quadratic term in Equation (11). Whether or not the quadratic term provides an appreciable enhancement to the collected charge depends on the size of the electronic pre-factor term $\frac{1}{2}KA_e v_e^s$ (recalling that for LGADs $A_{e/h}$ is appreciable only for electrons), which is the effective temporal scale over which the impact ionization process develops. Thus, the question of whether gain provides a benefit in the detection of X-ray fluxes depends upon the size of this pre-factor, which is set by the physics of the impact ionization process, and on the time scale over which charge is collected. The next section describes the results of a simulation study geared towards understanding the scale at which LGAD gain provides an advantage, relative to conventional PIN diodes, in X-ray detection.

## 3. Elemental Simulation

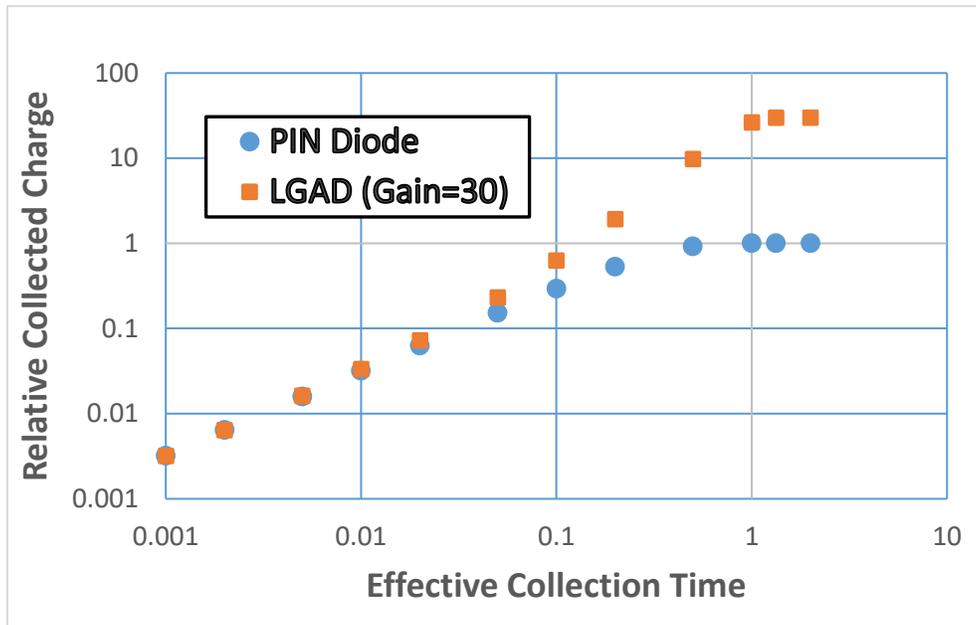

Figure 1: Simulated relative charge collection, as a fraction of charge deposited by the instantaneous X-ray flux, as a function of effective collection time, for PIN and LGAD sensors of 50 µm thickness. The relative collected charge can be greater than unity for the LGAD sensor due to the effect of impact ionization, which is tuned in the simulation to provide an overall gain of 30. Other simulated properties assumed for the sensors and their operation are described in further detail in the text.

To develop a feel for when LGADs might allow for a greater frame rate than conventional PIN diode sensors, Figure 1 provides a comparison of the collected charge as a function of effective collection time (related to the electronic shaping time, as discussed above, and in turn, inversely related to the achievable frame rate) between LGADs and PIN diodes. These results derive from a simulation of 50 µm thick planar (un-segmented) PIN and LGAD silicon diode sensors, with electron and hole drift speeds of 100 and 60 microns per nsec, respectively, taken from [6] under the assumption of room-temperature operation with a uniform bulk field of $2\times10^4$ V/cm. The simulation of the LGAD sensor includes, in addition, a 2 µm thick gain layer just below the positive (cathode) electrode in which impact ionization produces a limited avalanche and thus signal gain, and makes use of all-orders (exponential) growth of gain-layer signal with time rather than the linear approximation of Equation (6). Electrons and holes, whether produced in the initial deposition or later via impact ionization, are propagated through the sensor bulk until they reach the anode (holes) or cathode (electronics) electrode. The impact ionization coefficient α is assumed to be zero (no multiplication) for holes and uniform through the 2 µm gain layer for electrons. The value of the electron impact ionization coefficient $\alpha_e$ was chosen to provide a mean free path of 0.61 µm between impact ionization events, resulting in an overall gain, at infinite collection time, of 30 for the simulated LGAD sensor. Such a value would be expected for a gain-layer field on the order of $3\times10^5$ V/cm [7], which is typical for LGAD devices.

For the particular configuration explored in the simulation, it is seen that the secondary process of impact ionization commences on a short enough time scale that significant enhancement of the collected charge via the LGAD gain is achieved for collection times as short as 0.1 nsec, corresponding to frame rates on the order of 10 GHz. It should be noted that this conclusion would not hold in the granular limit, in that the arrival time of a signal produced by a single X-ray deposition would arrive at the gain layer over a range elapsed times of between 0 and 0.5 psec, leading to a corresponding variability in the onset of the bulk of the signal distribution, and limiting the frame rate to less than 2 GHz. The time development for a PIN sensor signal would be more uniform, although the size of the signal might be too small for consistent detection, again leading to a preference for an LGAD sensor.

## 4. Summary and Conclusions

An approximate, closed-form expression has been developed for the short time-frame behavior of silicon diode sensor signals in the context of high frame-rate detection of incident X-ray fluxes, in the limit that the X-ray absorption profile generates a longitudinally-uniform distribution of electron-hole pairs in the detector bulk. The expression represents the immediate time development of signals from diode sensors both with (LGAD) and without (PIN) gain. Principles limiting the detection frame rate in the presence of electronic readout noise are discussed, and it is seen that there is an advantage associated with signal gain, particularly as the desired frame rate increases. Making use of an elemental simulation, the relative advantage of LGAD vs. PIN diode sensors is explored as a function of the effective electronic collection time. It is found that for an idealized LGAD sensor with a gain of 30, the gain provided by impact ionization yields an advantage relative to PIN diode sensors for frame rates as high as 10 GHz.

**Acknowledgements**

This work was supported by the U.S. Department of Energy grant number DE-SC0010107, and the UC-National Laboratory Fees Research Program grant ID #LFR-20-653232.